\documentclass[]{spie}  %>>> use for US letter paper
\usepackage[]{graphicx}
\usepackage{amsmath,amssymb,mathtools,url}

\title{Characterization of microdot apodizers for imaging exoplanets with next-generation space telescopes} 
\usepackage{array}
\newcolumntype{P}[1]{>{\centering\arraybackslash}p{#1}}

\author{Manxuan Zhang\supit{a},  Garreth~Ruane\supit{a,$\dagger$}, Jacques-Robert~Delorme\supit{a}, Dimitri~Mawet\supit{a,b}, Nemanja~Jovanavic\supit{a}, Jeffrey~Jewell\supit{b}, Stuart~Shaklan\supit{b}, and J.~Kent~Wallace\supit{b}
\skiplinehalf
\supit{a}Department of Astronomy, California Institute of Technology, 1200 E. California Blvd.,\\Pasadena, CA 91125, USA; \\
\supit{b}Jet Propulsion Laboratory, California Institute of Technology, 4800 Oak Grove Dr.,\\Pasadena, CA 91109, USA
}

\authorinfo{$^\dagger$Send correspondence to gruane@caltech.edu}

\graphicspath{{Figures/}}
\begin{document} 
  \maketitle 
    \pagenumbering{arabic}
%%%%%%%%%%%%%%%%%%%%%%%%%%%%%%%%%%%%%%%%%%%%%%%%%%%%%%%%%%%%% 

\begin{abstract}
A major science goal of future, large-aperture, optical space telescopes is to directly image and spectroscopically analyze reflected light from potentially habitable exoplanets. To accomplish this, the optical system must suppress diffracted light from the star to reveal point sources approximately ten orders of magnitude fainter than the host star at small angular separation. Coronagraphs with microdot apodizers achieve the theoretical performance needed to image Earth-like planets with a range of possible telescope designs, including those with obscured and segmented pupils. A test microdot apodizer with various bulk patterns (step functions, gradients, and sinusoids) and 4 different dot sizes (3~$\mu$m, 5~$\mu$m, 7~$\mu$m, and 10~$\mu$m) made of small chrome squares on anti-reflective glass was characterized with microscopy, optical laser interferometry, as well as transmission and reflectance measurements at wavelengths $\lambda$=600~nm and $\lambda$=800~nm. Microscopy revealed the microdots were fabricated to high precision. Results from laser interferometry showed that the phase shifts observed in reflection vary with the local microdot fill factor. This effect is not explained purely by interference between reflected fields from the chrome and glass portions. %Although such interference is not likely to dominate our measurements, it may be mitigated by maximizing the reflectively ratio between the chrome and exposed glass regions and the tuning the height of the dots to be $\lambda$/2. 
Transmission measurements showed that microdot fill factor and transmission were linearly related for dot sizes $\ge$5~$\mu$m. However, anomalously high transmittance was measured when the dot size is $<$5$\times$ the wavelength (i.e. $\sim$3~$\mu$m) and the fill factor is approximately 50\%, where the microdot pattern becomes periodic. The transmission excess is not as prominent in the case of larger dot sizes suggesting that it is likely to be caused by the interaction between the incident field and electronic resonances in the surface of the metallic microdots. We used our empirical models of the microdot apodizers to optimize a second generation of reflective apodizer designs, specifically for demonstrating end-to-end instrumentation for planet characterization at Caltech's High Contrast Spectroscopy Testbed for Segmented Telescopes (HCST), and confirmed that the amplitude and phase of the reflected beam closely matches the ideal wavefront. 
\end{abstract}

%>>>> Include a list of keywords after the abstract 


%%%%%%%%%%%%%%%%%%%%%%%%%%%%%%%%%%%%%%%%%%%%%%%%%%%%%%%%%%%%%
\section{INTRODUCTION}
\label{sec:intro}  % \label{} allows reference to this section

Direct imaging and spectroscopy of Earth-like exoplanets orbiting Sun-like stars with future space telescopes requires an optical system that sufficiently suppresses light from the host star to detect planets that are $<10^{-10}$ times fainter, at small angular separations (a few tens of milliarcseconds), over a broad spectral range ($\Delta\lambda/\lambda\gtrsim0.1$), while preserving the faint planet signal. These goals are critical to the science case of two mission concepts currently under study by NASA in preparation for the 2020 Astrophysics Decadal Survey: the Habitable Exoplanet Imaging Mission (HabEx)\cite{Mennesson2016} and the Large UV/Optical/IR Surveyor (LUVOIR)\cite{Pueyo2017}. The HabEx and LUVOIR mission concepts are ultra-stable space telescopes with primary mirror diameters of 4-15~m and a range of possible architectures including a 4~m off-axis, monolithic telescope or a much larger ($>$6~m) segmented telescope. Detecting biosignatures on potentially habitable planets is a premier science driver of each mission concept. 

Apodization is used in exoplanet imaging to reduce the amount of diffracted starlight at small angular separations from the host star where planets may be detected \cite{Watson1991,Nisenson2001,Kasdin2003}. Originally developed to maximize the sensitivity of ground-based adaptive instruments by reducing systematic speckle noise \cite{Angel1994,Sivaramakrishnan2002,Perrin2003,Soummer2007}, apodized coronagraphs make use of masks in additional planes to further improve the rejection of starlight rejection. Over a decade of research has made significant progress in identifying high-performance pupil apodization patterns for conventional Lyot \cite{Soummer2003_APLC,Soummer2005_APLC,Soummer2009_APLCII,Soummer2011_APLCIII,NDiaye2015_APLCIV,NDiaye2016_APLCV} and phase mask coronagraphs\cite{Soummer2003_DZPM,Guyon2014,Mawet2013_ringapod,Ruane2016_SPIE,Jewell2017,Ruane2018_JATIS,Fogarty2017}. Many of these designs call for an achromatic, gray-scale field amplitude in the entrance pupil that minimizes the power at mid-spatial frequencies after the coronagraph. 

Microdot apodizers approximate the desired gray scale pupil function by patterning small metallic dots on a glass substrate whose local density is matched to the prescribed transmission or reflection. This approach has been used for beam shaping of high powered lasers\cite{Dorrer2007} as well as in the apodized Lyot coronagraphs of two of the most prolific ground-based exoplanet imaging instruments: the Gemini Planet Imager (GPI)\cite{Macintosh2006} and the Spectro-Polarimetric High-contrast Exoplanet REsearch instrument (SPHERE)\cite{SPHERE2008}. Extending the use of microdot apodizers to space-based exoplanet imaging may require more complicated apodization patterns for suppressing the diffracted light from spider support structures and gaps between mirror segments. In addition, the apodizer will likely be used in reflection to avoid chromatic wavefront aberrations introduced by the glass substrate. 

Experiments performed here expand on previous efforts to empirically characterize microdot apodizers in transmission for ground-based infrared imaging instruments\cite{Martinez2009a,Martinez2009b,Sivaramakrishnan2009,Thomas2011,Mawet2014} by including interferometric and reflectance measurements to validate models of the reflected wavefront and inform the design of coronagraphs for future space telescopes. Microdot test patterns based on square microdots in chrome deposited on anti-reflection (AR) coated glass were designed, manufactured, and characterized. The test sample represented a range of possible amplitude functions and dot sizes (3-10 $\mu$m). An analytical model of the reflected wavefront calibrated by our experiments was then used to fabricate a gold microdot apodizer for a segmented aperture vortex coronagraph to be tested on Caltech's High Contrast Spectroscopy Testbed for Segmented Telescopes (HCST)\cite{Delorme2017}.

\section{Characterization of the Microdot Apodizer Test Sample} 

\begin{figure}[t]
    \centering
    \includegraphics[height=5.0cm,trim={0 -2cm 0 0},clip]{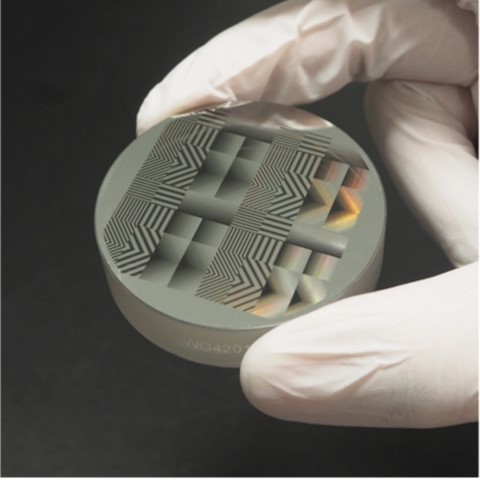}
    \includegraphics[height=5.3cm,trim={0 0 2.5cm 0},clip]{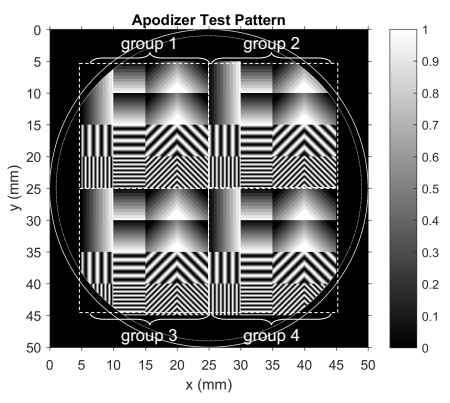}
    \includegraphics[height=5.1cm]{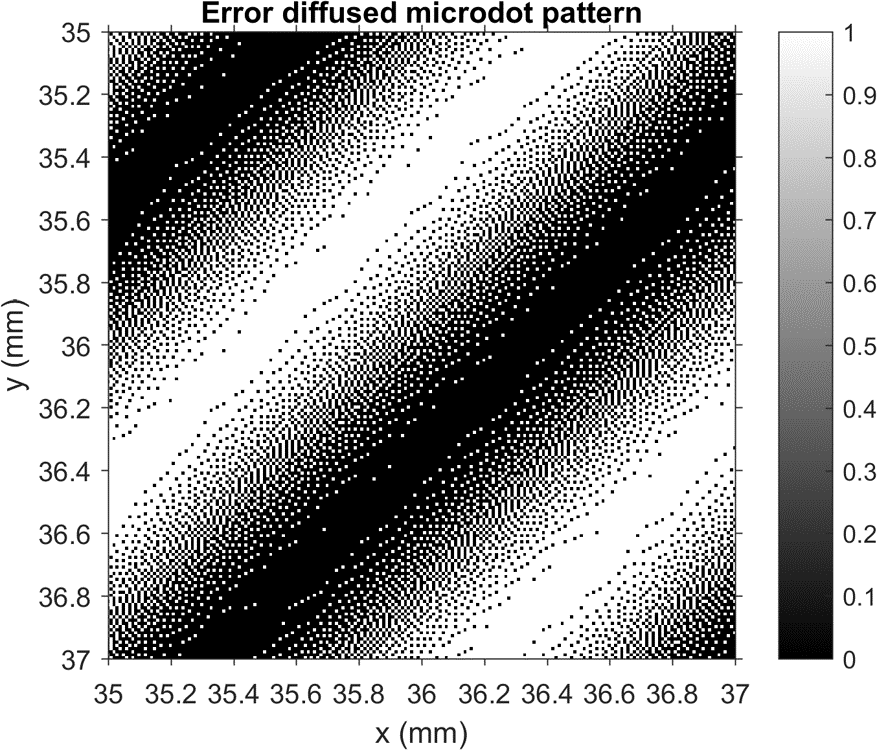}
    \caption{Apodizer test pattern manufactured with a chrome layer deposited on an anti-reflection-coated glass substrate with a diameter of $\sim$50~mm. 
    \textit{Left:} A photograph of the sample. \textit{Middle:} The designed transmission pattern. Groups 1-4 have consist of 16 square pattern with dot sizes of 3~$\mu$m, 5~$\mu$m, 7~$\mu$m, and 10~$\mu$m, respectively. \textit{Right:} Zoom in on a sinusoidal pattern in group 4 showing the specified error-diffused microdot pattern (not an actual image).}
    \label{fig:ApodizerTestSample}
\end{figure}

\begin{figure}[t]
    \centering
    \includegraphics[width=0.6\linewidth]{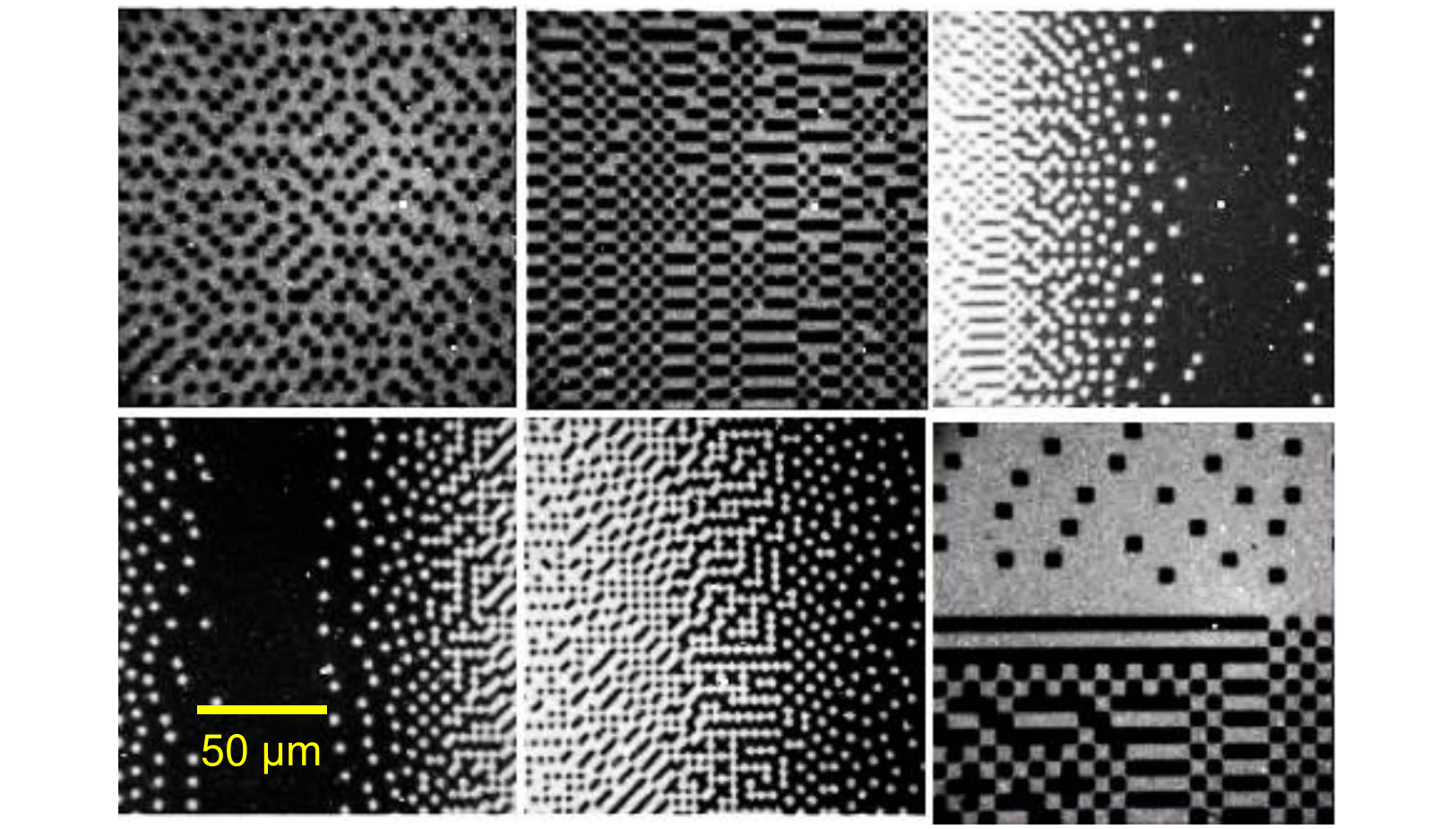}
    \caption{Optical microscope images of various regions of the test sample confirmed that the manufacturer reproduced the designed patterns to within $\sim$1~$\mu$m.}
    \label{fig:microscopy}
\end{figure}

An apodizer test sample was fabricated to empirically characterize the wavefront reflected from a metallic microdots of various density and size. The microdots were deposited in a 113.5~nm thick layer of chrome with a reflectance of 53\% at $\lambda$=600~nm on a fused silica AR-coated substrate (Thorlabs WG42012-A) by Opto-Line International, Inc. The test pattern was made up of 64 different 5~mm~$\times$~5~mm squares with a variety of spatially-varying reflectance patterns including sinusoids, gradients, or step functions (see Fig. \ref{fig:ApodizerTestSample}). A set of 16 patterns was replicated 4 times with dot sizes of 3~$\mu$m, 5~$\mu$m, 7~$\mu$m, and 10~$\mu$m. The semi-random distribution of dots was determined from continuous input functions using a Floyd-Steinberg dithering algorithm\cite{FloydSteinberg1976,Ulichney1988}, also known as error diffusion or halftoning. Optical microscope images confirmed that the manufacturer reproduced the designed pattern to within $\sim$1~$\mu$m spatial precision (see Fig. \ref{fig:microscopy}). In the following, we characterize reflected and transmitted wavefront from the test sample in order to empirically determine the effect of the metallic layer thickness and dot size.

% An apodizer test sample was made of chrome on a glass substrate with 64 different 5 mm x 5 mm squares with different patterns (either sinusoids, gradients, or step functions) and microdot sizes (3 μm, 5 μm, 7 μm, 10 μm) was made with the intent of experimentally determining the transmittance and reflectance of grayscale apodizer masks. The accuracy of the phases and amplitudes along with the anomalies will be helpful in deciding on materials, patterns, microdot sizes for future apodizers. 

% \subsection{Microscopy}

% To ensure that the manufacturer had produced an apodizer of high quality, I took images at 27.5 times magnification and showed that the test sample was of high craftsmanship. The microdots were shown to be perfectly square in 11 different regions and at two different magnifications.

\subsection{Laser interferometry}

\begin{figure}[t]
    \centering
    \includegraphics[height=5.05cm]{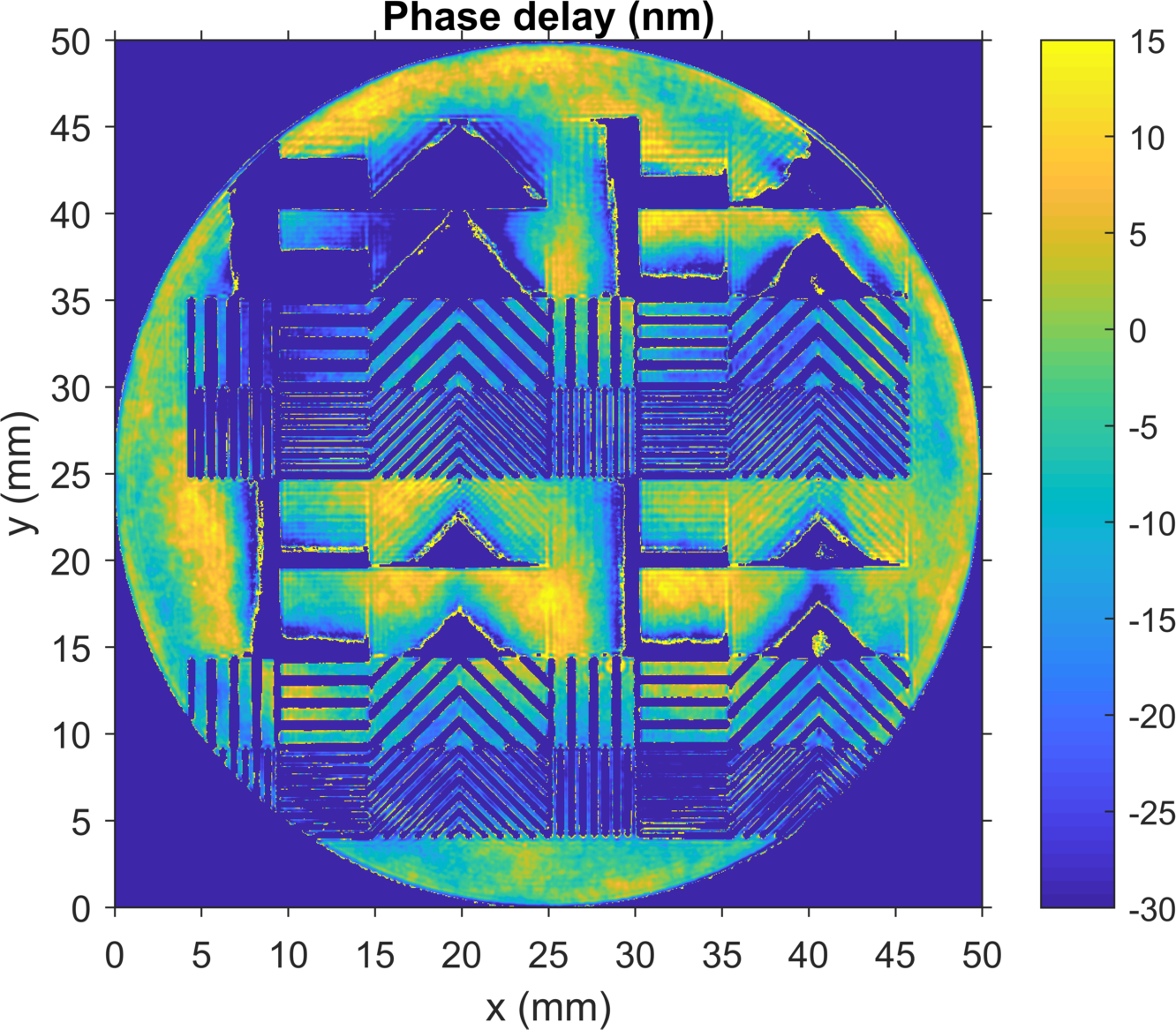}
    \includegraphics[height=4.9cm]{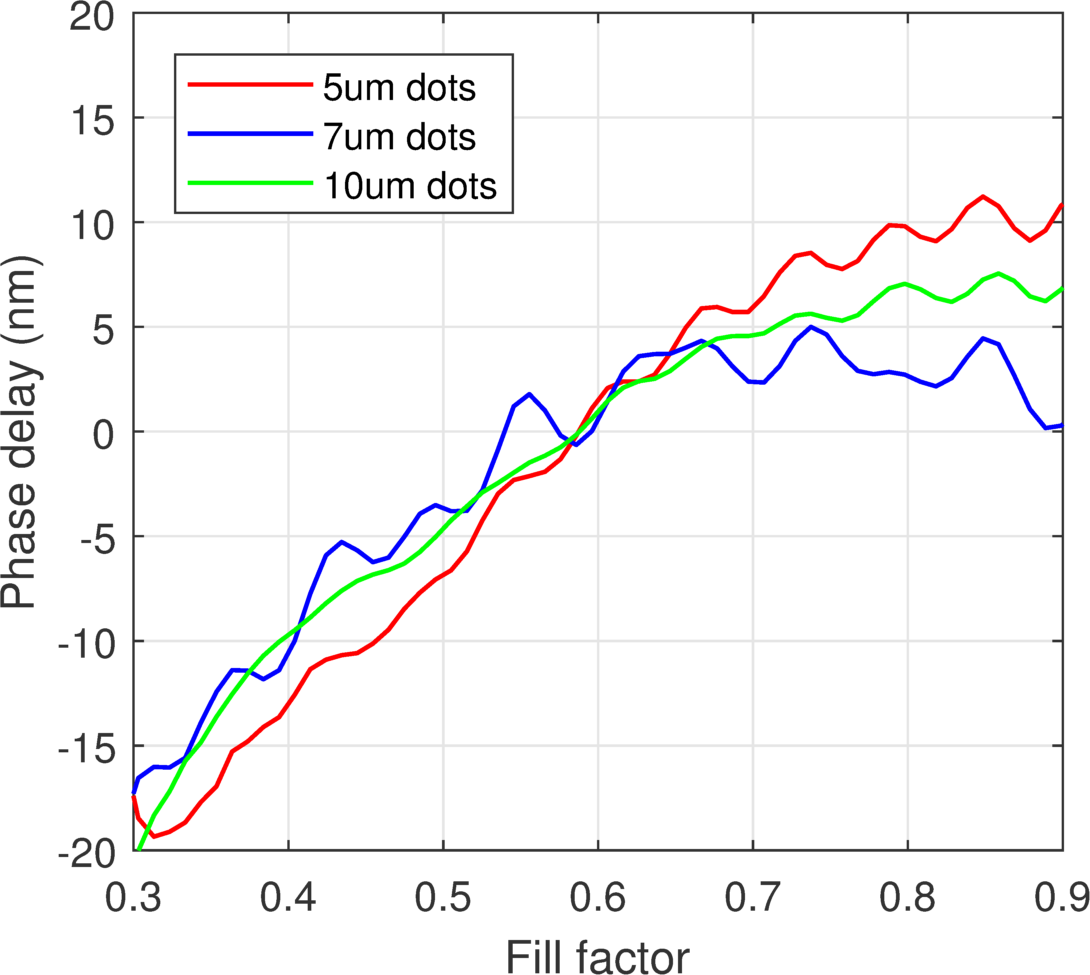}
    \includegraphics[height=4.9cm]{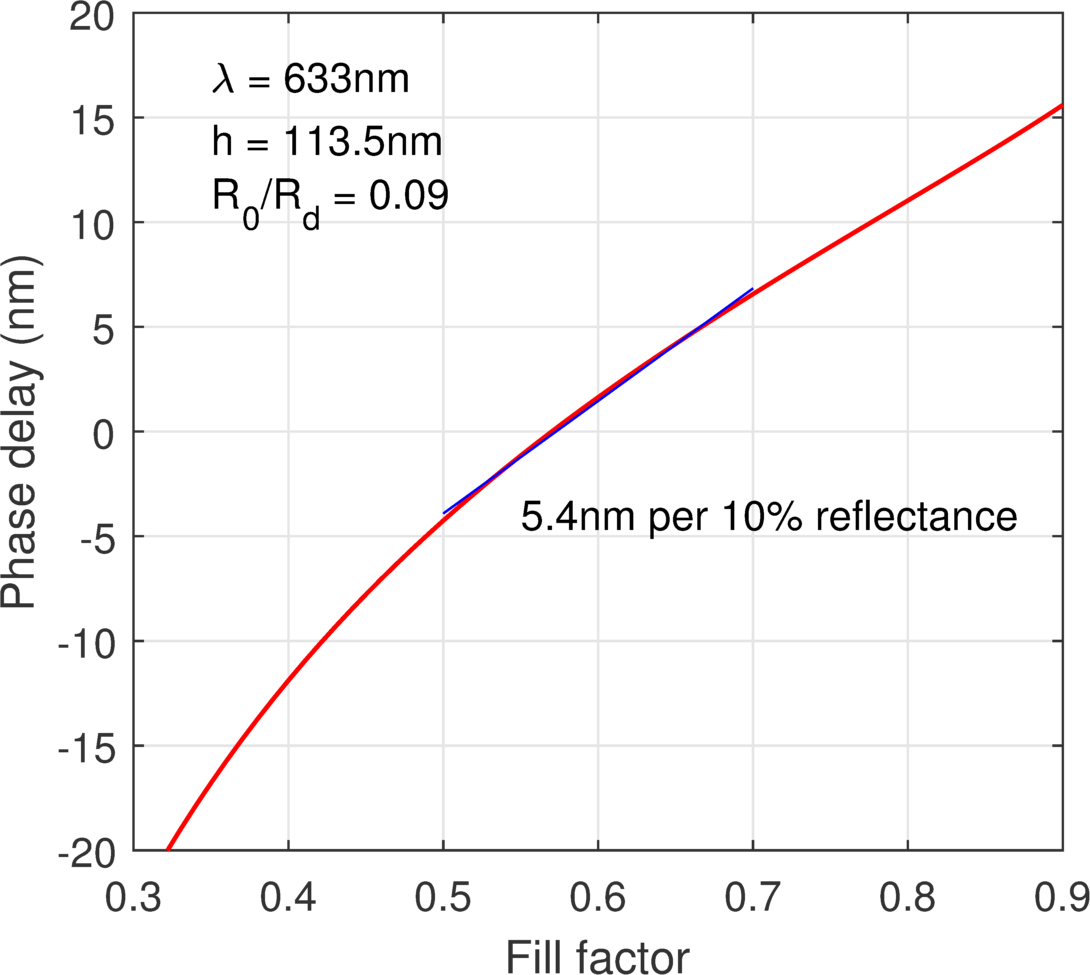}
    \caption{\textit{Left:} Phase delay (nm) upon reflection at $\lambda$=633~nm measured using a Fizeau laser interferometer with low order aberrations removed. The ramps and step functions in reflection show visually that the phase depends on the local microdot fill factor. Data was not obtained in regions of very low reflectance and is missing in the image. \textit{Middle:}  Line profiles along paths of linearly increasing fill factor. \textit{Right:} Phase delay expected from the OPD interference model at $\lambda$=633~nm, a layer thickness of $h$=113.5~nm, and reflectance ratio of $R_0/R_d$=0.09.}
    \label{fig:interfero}
\end{figure}

The phase of the beam reflected from the sample at $\lambda$=633~nm was measured using a Fizeau laser interferometer (Zygo Verifire). The measured phase delay as a function of position was dominated by low order aberrations $>$100~nm peak-to-valley. Although the substrate was specified to have a surface flatness of $\lambda/10$, additional spherical aberration was introduced by our mounting scheme. Removing the low order aberrations by projecting the measured phase onto the Zernike basis and subtracting the reconstructed wavefront revealed phase structures which are correlated with the local microdot fill factor  (Fig. \ref{fig:interfero}, \textit{left}). However, a flat reflected wavefront is generally desired for apodizers in coronagraphs. 

Phase delays measured along paths with linearly increasing fill factor were extracted from several squares on the sample (Fig. \ref{fig:interfero}, \textit{middle}). Within the range of fill factors from 0.4 to 0.7, the phase delay appears to be roughly linear for 5-10~$\mu$m dots with an average phase delay of $\sim$5.4~nm per 10\% change in reflectance. Similar phase effects have been previously identified in microdot apodizers by Mawet et al. (2014)\cite{Mawet2014}. Regions where the fill factor is $<$0.4 were not well characterized by the interferometer because the reflected signal becomes too faint. Residual low order surface errors dominated in the regions with 3~$\mu$m dots. The phase delay tended to deviate significantly from the linear dependence at fill factors $>$0.7. 

\subsection{Analytical modeling of the reflected wavefront}

We set out to develop an analytical, superposition model to explain the measured phase effects. In this scheme, the effective scalar field reflected from a microdot apodized mask may be approximated as the spatial mean of the fields reflected over a region much larger than the dot size: $E_\mathrm{eff}  = \langle E_\mathrm{d} + E_\mathrm{0}  \rangle$, where $E_\mathrm{d}$ and $E_\mathrm{0}$ are the reflected fields from the dots and substrate, respectively. For microdots with reflectance $R_d=r_d^2$ and fill factor $f_d$ deposited on a substrate of reflectance $R_0=r_0^2$, the spatial average of the fields from each region is 
 \begin{equation}
 	E_\mathrm{eff} = r_d \sqrt{f_d}  e^{i \Delta \phi} + r_0 \sqrt{1-f_d} ,
 	\label{eqn:model}
 \end{equation}
where $\Delta \phi$ is the relative phase shift between regions reflected from the metal and substrate. We hypothesized that the phase shift could be explained by the optical path difference (OPD) between the metal and substrate; that is, $\Delta \phi=4\pi h / \lambda$ and $h$ is the thickness of the metallic layer ($\sim$113.5~nm). The phase of $E_\mathrm{eff}$ is plotted in Fig. \ref{fig:interfero}, \textit{right} for a set of parameters consistent with the measured phase. The best fit reflectance ratio $R_0/R_d\approx0.09$ in this simple approximation is significantly higher than expected. The specification of the AR-coated substrate and chrome material indicate a ratio of $R_0/R_d$=0.0013/0.53=0.0025, which would yield a phase delay $<$1~nm instead of the measured $\sim$5~nm per 10\% change in fill factor. The effective $R_0/R_d$ ratio is approximately 36 times higher than expected. There are several potential explanations for this difference. (1) It could be due to damage to the AR coating during the deposition of the chrome layer. Without the AR coating, the reflectance of the substrate is $\sim$4\%. (2) Another potential source of error arises from features in the microdot pattern, especially corners, which are small enough that our scalar wave optical model of the reflected wavefront breaks down, leading to lower effective $R_d$. The model agrees with the measurements if, for example, $R_0\approx0.04$ and $R_d\approx40\%$.

The discrepancy between the measurements may be due to the interaction between the incident field and electronic resonances in metal films, which is not described by the theoretical description above. Such effects may play a significant role in the reflectance and transmission of the microdot apodizer samples \cite{Ebbesen1998,Genet2007}. The experiments performed here are intended the calibrate the analytical model empirically. However, as explained in the next section, we found that the phase delay does not follow the expected $1/\lambda$ trend and the OPD model calibrated at one wavelength can not generally be applied to another.

\subsection{Reflectance measurements: comparison with analytical models}

\begin{figure}[t]
    \centering
    \includegraphics[width=0.5\linewidth]{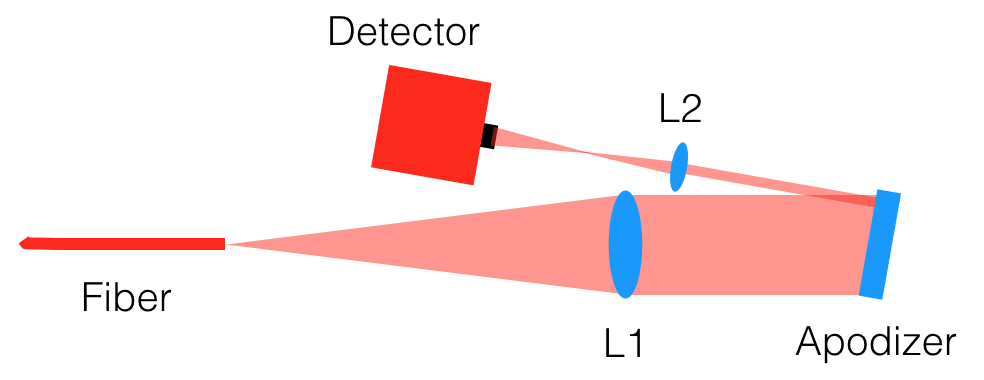}
    \caption{Experimental setup to measure the reflectance of the apodizer test samples. Light from a fiber-coupled super-continuum source was collimated by lens L1 with a focal length of 200~mm. The beam diameter was $\sim$25~mm, overfilling the 5~mm~$\times$~5~mm squares on the test sample, at an angle of incidence on the sample was 8$^\circ$. Lens L2 (focal length of 100~mm) re-imaged the reflected beam onto a 6.7~mm~$\times$~5.3~mm detector with 1:1 magnification.}
    \label{fig:refl_setup}
\end{figure}

The reflectance of the sample was measured at $\lambda$=600$\pm$5~nm (Thorlabs FB600-10 filter) and $\lambda$=800$\pm$5~nm (Thorlabs FB800-10 filter) using the optical layout illustrated in Fig. \ref{fig:refl_setup}. The sample was illuminated with a collimated beam from a fiber-coupled super-continuum source (NKT Photonics SuperK) at an angle of incidence of 8$^\circ$. The beam width at the apodizer was 25~mm, 5 times larger than the 5~mm~$\times$~5~mm squares on the test sample. The reflected beam was re-imaged with 1:1 magnification on to a 6.7~mm~$\times$~5.3~mm CMOS detector (Thorlabs DCC1545M; 1280$\times$1024 pixels). All 64 squares were imaged individually.

\begin{figure}[t]
    \centering
    \includegraphics[height=5.0cm]{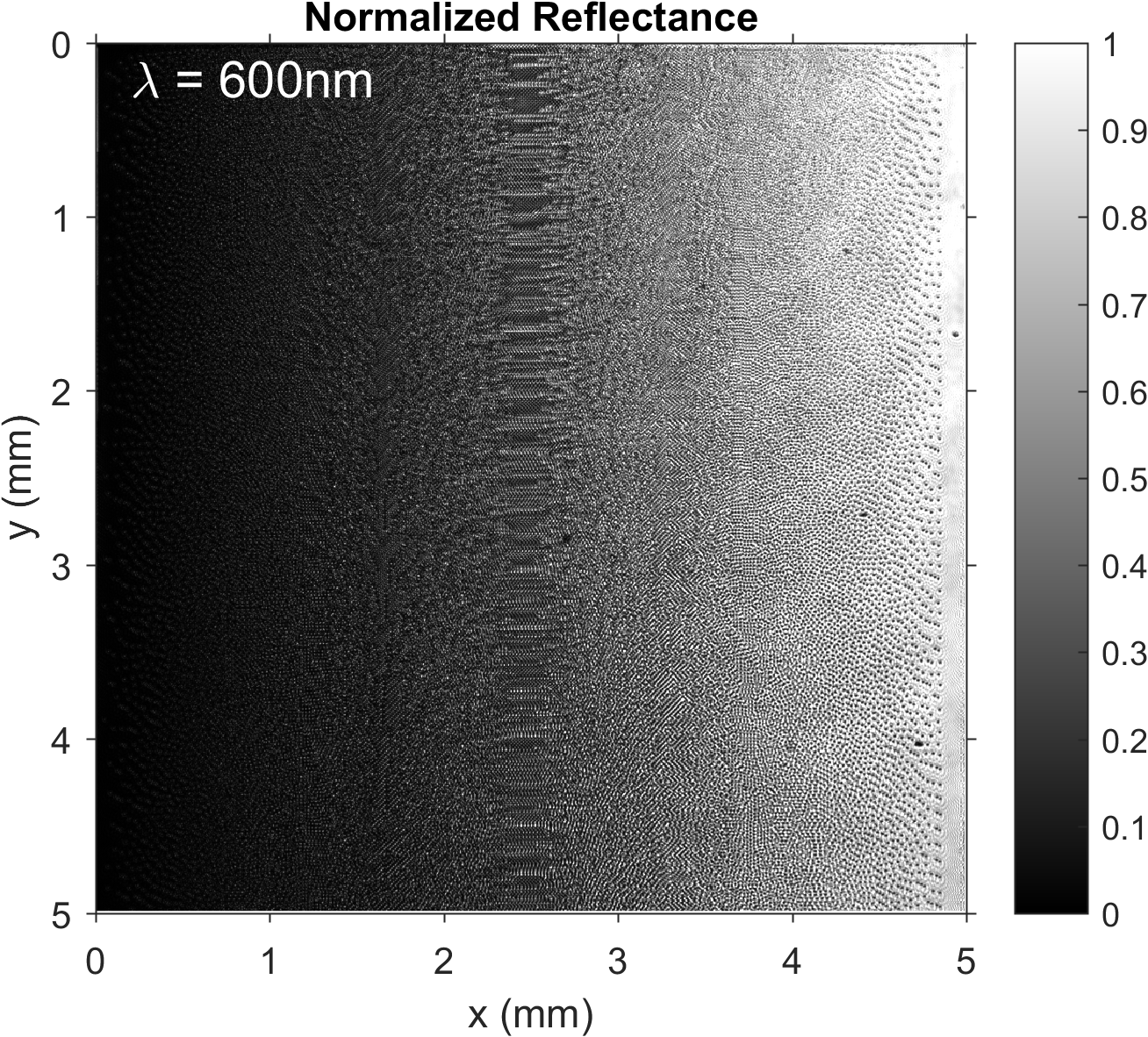}\hspace{2mm}
    \includegraphics[height=4.9cm]{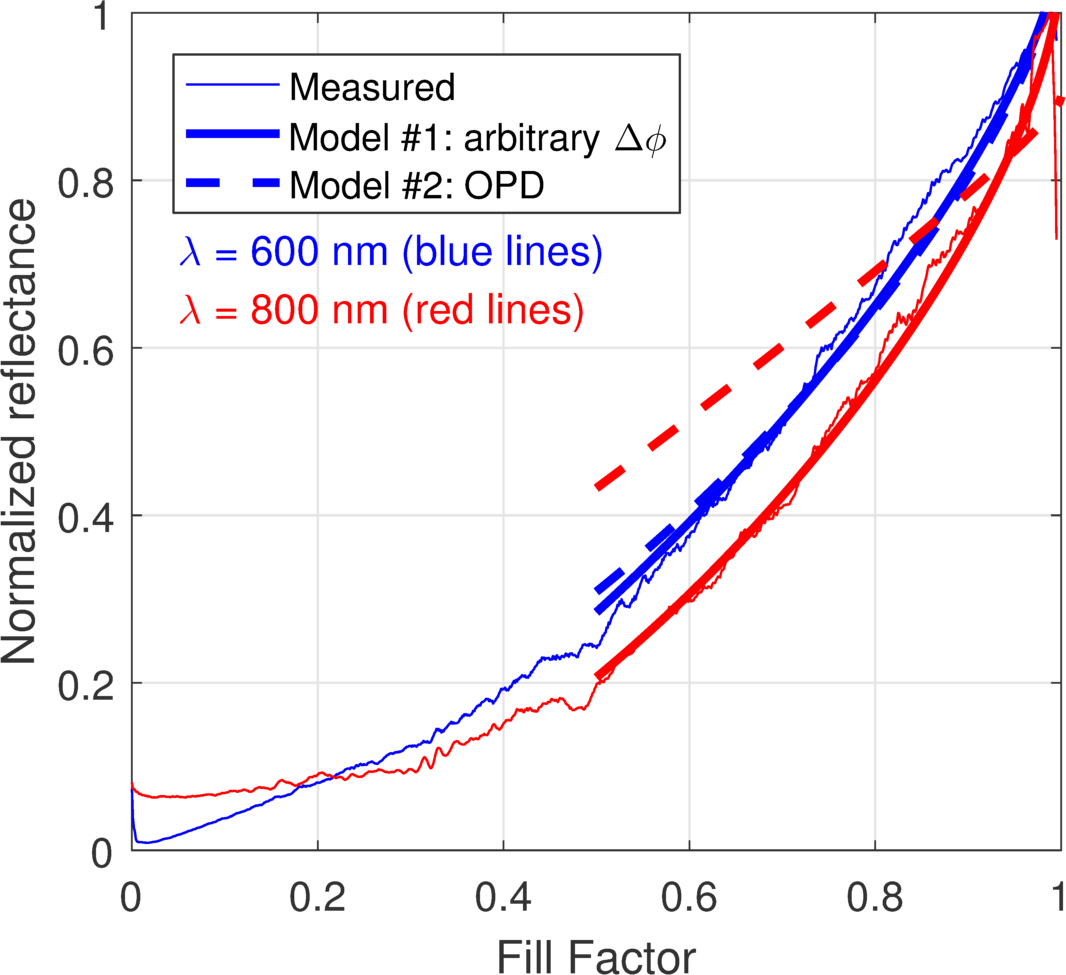}
    \caption{\textit{Left:} Example image of a 5~mm~$\times$~5~mm square with a linearly increasing fill factor of 10~$\mu$m dots. \textit{Right:} The median measured horizontal line profile along the vertical axis (thin solid lines) for the 600~nm (blue) and 800~nm (red) filters along with two model fits. The first (model \#1, thick solid lines) allows the phase delay to take on an arbitrary value as a function of wavelength whereas the second (model \#2, dashed lines) assumes the phase delay expected from the OPD assuming $h$=113.5~nm. The second model does not agree well the measured reflectance as a function of wavelength. }
    \label{fig:refl_modelfits}
\end{figure}

Figure 5, \textit{left} shows an example image at $\lambda$=600$\pm$5~nm of a square with a linearly increasing microdot fill factor from group 4 in Fig. \ref{fig:ApodizerTestSample}. We computed the median horizontal line profile (see Fig. 5, \textit{right}, thin blue line) and fit two models based on the following expression for the normalized reflectance 
% \begin{equation}
%     R =& \left| \sqrt{f_d} r_d e^{i \Delta \phi} + \sqrt{1-f_d} r_0 \right|^2
% \end{equation}
% \begin{equation}
%     \frac{R}{R_d} = f_d + (1-f_d)\frac{R_0}{R_d}+2\sqrt{f_d(1-f_d)\frac{R_0}{R_d}}\cos(\Delta\phi).
% \end{equation}
\begin{equation}
    R/R_d = f_d + (1-f_d)(R_0/R_d)+2\sqrt{f_d(1-f_d)(R_0/R_d)}\cos(\Delta\phi).
\end{equation}
In the first fit (model \#1, see Fig. 5, \textit{right}, thick blue line), we used three free parameters: the effective reflectance ratio between the metal and glass ($R_0/R_d$), the relative phase delay ($\Delta\phi$), and an overall reflectance normalization factor. In the second (model \#2, see Fig. 5, \textit{right}, dashed blue line), we fixed the phase shift to $\Delta \phi=4\pi h / \lambda$, where $h$=113.5~nm. We then repeated the measurement and analysis for $\lambda$=800$\pm$5~nm (see Fig. 5, \textit{right}, red lines). The best fit parameters are listed in Table \ref{tab:fits}.

\begin{table}[h]
\caption{Best fit parameters calculated for the models shown in Fig. \ref{fig:refl_modelfits}.}
\label{tab:fits}
    \centering
    \begin{tabular}{|l|c|c|c|c|}
        \cline{2-5}
        \multicolumn{1}{c|}{ }& \multicolumn{2}{c|}{$\lambda$=600~nm} & \multicolumn{2}{c|}{$\lambda$=800~nm}\\
        \hline
        Model \# & $R_0/R_d$ & $\Delta\phi$ & $R_0/R_d$ & $\Delta\phi$\\
        \hline
        1 (arbitrary $\Delta\phi$) & 0.11 & 1.1$\pi$ & 0.20 & 1.2$\pi$ \\
        \hline
        2 (OPD, $h$=113.5~nm) & 0.14 & 0.76$\pi$ & 0.04 & 0.57$\pi$ \\
        \hline
    \end{tabular}
\end{table}

We found that model \#1 accurately explains the measured reflectance at each wavelength individually. The best fit relative reflectances ($R_0/R_d$) were 0.11 at $\lambda$=600~nm and 0.2 at $\lambda$=800~nm. Although the reflectance of chrome is expected to be achromatic over this wavelength range, the factor of $\sim$2 change in relative reflectance may be due to the AR coating on the substrate which is only designed for 400-700~nm (Thorlabs coating `A') assuming it survived the manufacturing process. The best fit for $\Delta\phi$ was $\sim\pi$ at both wavelengths. On the other hand, OPD model (\#2) did not accurately predict the measured reflectance and a function of wavelength. The best fit for $R_0/R_d$ at $\lambda$=600~nm was consistent with that of model \#1 as well as the value predicted from the interferometeric measurements at $\lambda$=633~nm. However, the best fit value of $R_0/R_d=0.04$ at $\lambda$=800~nm for the OPD model (\#2) is not realistic and nevertheless there was large discrepancy between the best fit model and the measured reflectance.  

We concluded that although the analytical expressions presented above were useful for parameterizing an empirically calibrated model to describe the amplitude and phase of a beam reflected from the microdot apodizer, they were generally insufficient for representing the underlying physics and predicting the reflected field as a function of wavelength. Our results implied that the true amplitude and phase of the reflected field is less dependent on wavelength compared to what one may expect from simple OPD models. It is probable that other effects dominate the measured amplitude and phase. Future work will apply finite difference time domain (FDTD) and rigorous coupled wave analysis (RCWA) techniques to model these effects and potentially help develop methods to mitigate them.

\subsection{Transmission measurements: the effect of dot size}

The measurements in reflection described above did not show a strong dependence on the size of the microdots from 5~$\mu$m to 10~$\mu$m. In this section, we use optical transmission measurements to show that anomalies do appear when the dot size approaches the wavelength. Since the transmission through the chrome material is very small ($\sim10^{-4}$), measuring the transmission of the sample as a function of fill factor may avoid phase effects from interference between the field reflected from the microdots and substrate. 

\begin{figure}[t]
    \centering
    \includegraphics[width=0.9\linewidth]{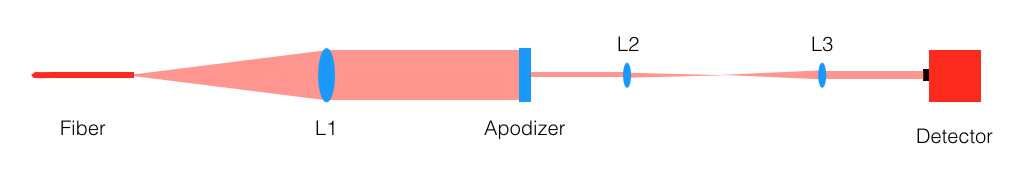}
    \caption{Experimental setup to measure the transmission of the sample. Light from a fiber-coupled super-continuum source was collimated by lens L1 with a focal length of 200~mm. The beam diameter was $\sim$25~mm, overfilling the 5~mm~$\times$~5~mm squares on the test sample. Lenses L2 and L3 (focal lengths of 100~mm) re-imaged the transmitted beam onto a 6.7~mm~$\times$~5.3~mm detector with 1:1 magnification.}
    \label{fig:trans_setup}
\end{figure}

Figure \ref{fig:trans_setup} shows the optical layout used to measure the transmission of the apodizer sample as a function of fill factor. Using the same source, filters, optics, and detector as in the reflectance measurements, we illuminated the sample at normal incidence and imaged the transmitted beam using an optical relay with 1:1 magnification and imaged each of the 64 squares individually. We extracted each square in post-processing and computed the median transmission profile along paths of linearly increasing fill factor. 

\begin{figure}[t]
    \centering
    \includegraphics[height=5.2cm]{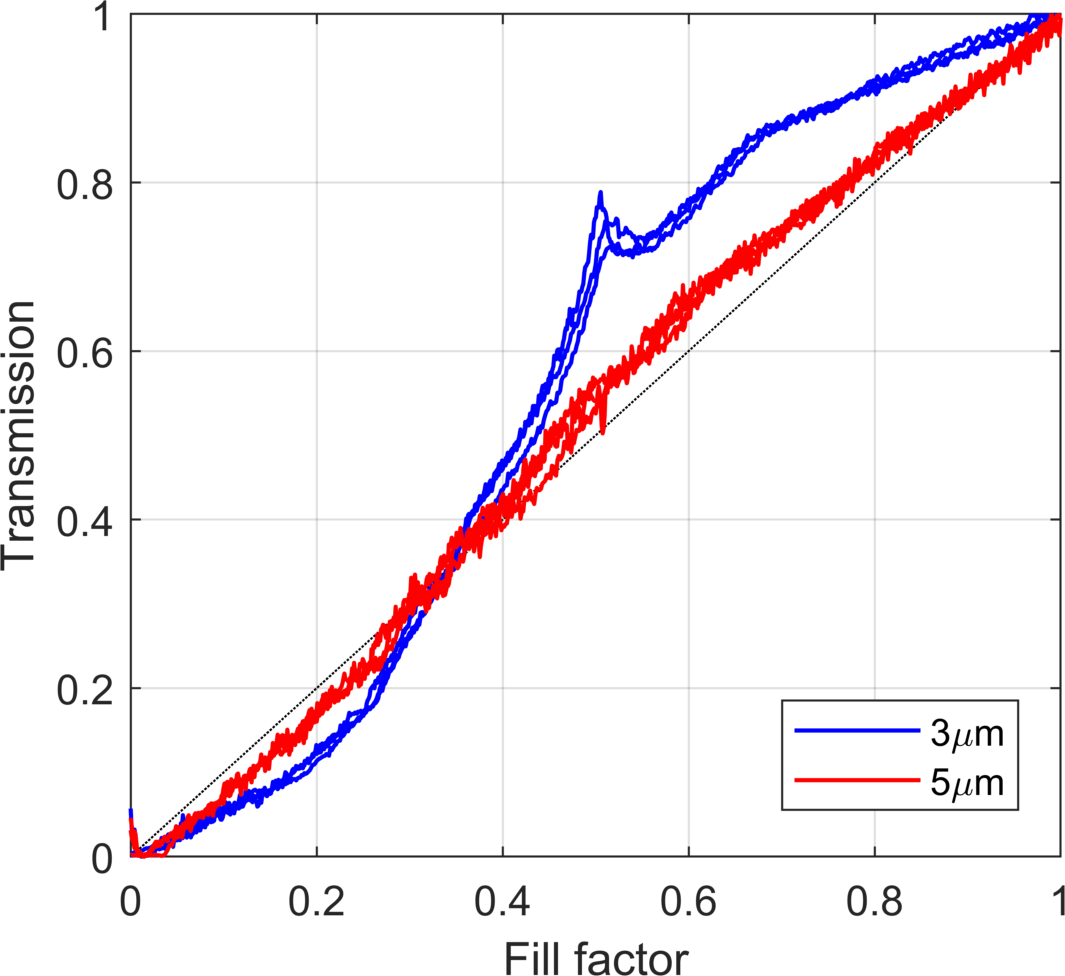}
    \includegraphics[height=5.3cm]{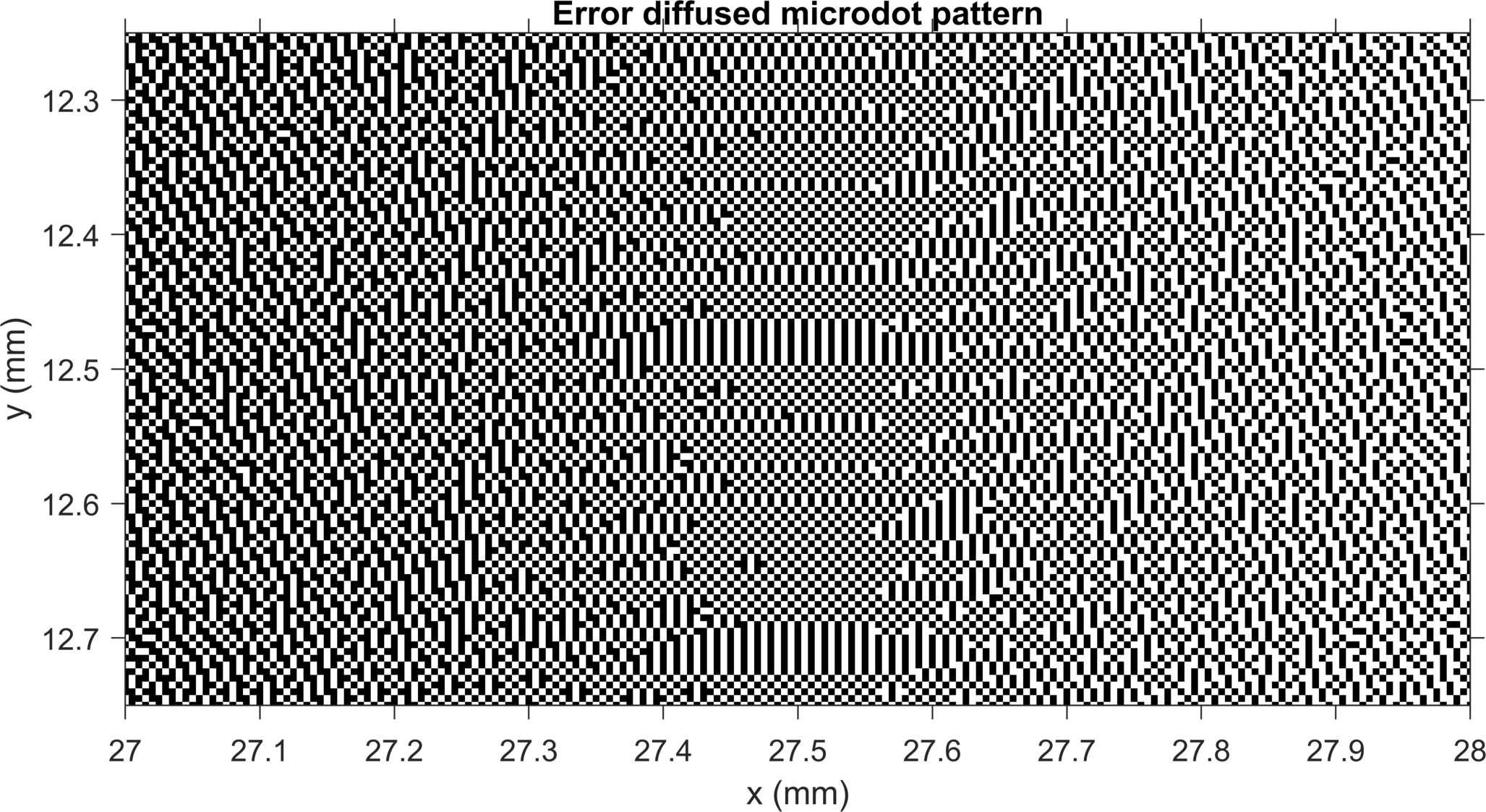}
    \caption{\textit{Left:} Median line profiles along paths of linearly increasing fill factor for three squares on the apodizer test sample with 3~$\mu$m and 5~$\mu$m microdots. \textit{Right:} Zoom in on the prescribed 5~$\mu$m microdot test pattern (see Fig. \ref{fig:ApodizerTestSample}) showing a range of fill factors from 0.4 to 0.6 and the appearance of a checkerboard pattern at 50\% fill factor, which causes anomalously high transmission.}
    \label{fig:trans_vs_dotsize}
\end{figure}

We found that the transmission as a function of fill factor becomes very nonlinear for the 3~$\mu$m dot size with enhanced transmission ($\sim$0.1-0.3 larger than expected) in areas where the fill factor is $\sim$50\% (see Fig. \ref{fig:trans_vs_dotsize}, \textit{left}). On the other hand, dot sizes of 5~$\mu$m appeared to have a roughly linear dependence with minor discrepancies (transmission error of $<$0.1). These observations were consistent with previous measurements made in the context of the GPI apodizer masks\cite{Sivaramakrishnan2009}. This phenomenon was found to be concurrent with the appearance of periodic patterns in the error diffused mask. Figure \ref{fig:trans_vs_dotsize}, \textit{right} shows a zoomed in image of the designed pattern. At 50\% fill factor, the microdots were arranged in a periodic checkerboard pattern, which achieved the desired fill factor with a grid of square microdots. The origin of the enhanced transmission may potentially be explained by the interaction between the incident field with electronic resonances in the surface of the metal film \cite{Ebbesen1998,Genet2007}. Using a relatively large dot size ($\sim$10$\times$) the wavelength and avoiding continuous regions of $\sim$50\% fill factor will naturally yield microdot apodizers that are well matched to the desired amplitude and phase of the transmitted or reflected wavefront. 

\begin{figure}[t!]
    \centering
    \includegraphics[height=5.45cm,trim={0.4cm 0.4cm 0 0},clip]{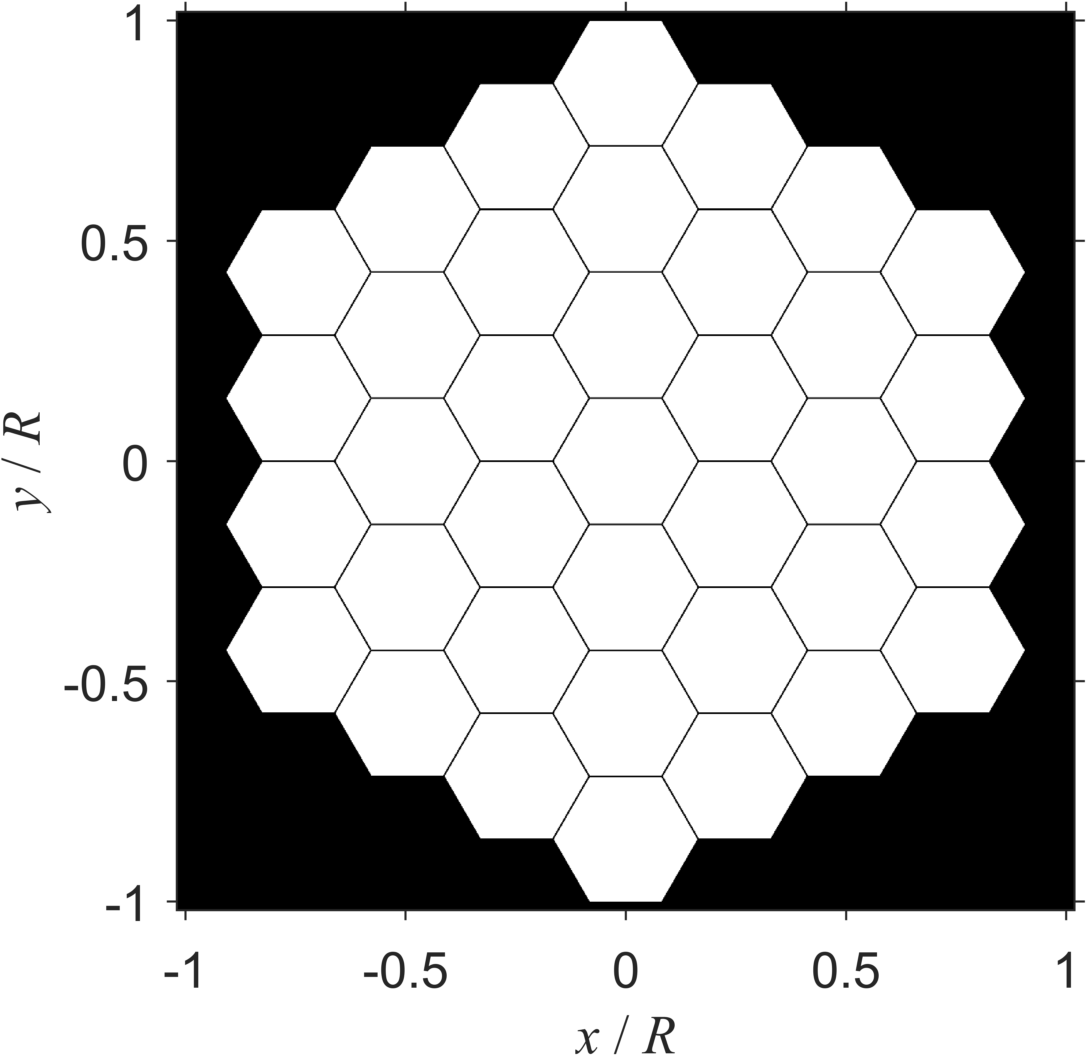}
    \includegraphics[height=5.45cm,trim={0.4cm 0.4cm 0 0},clip]{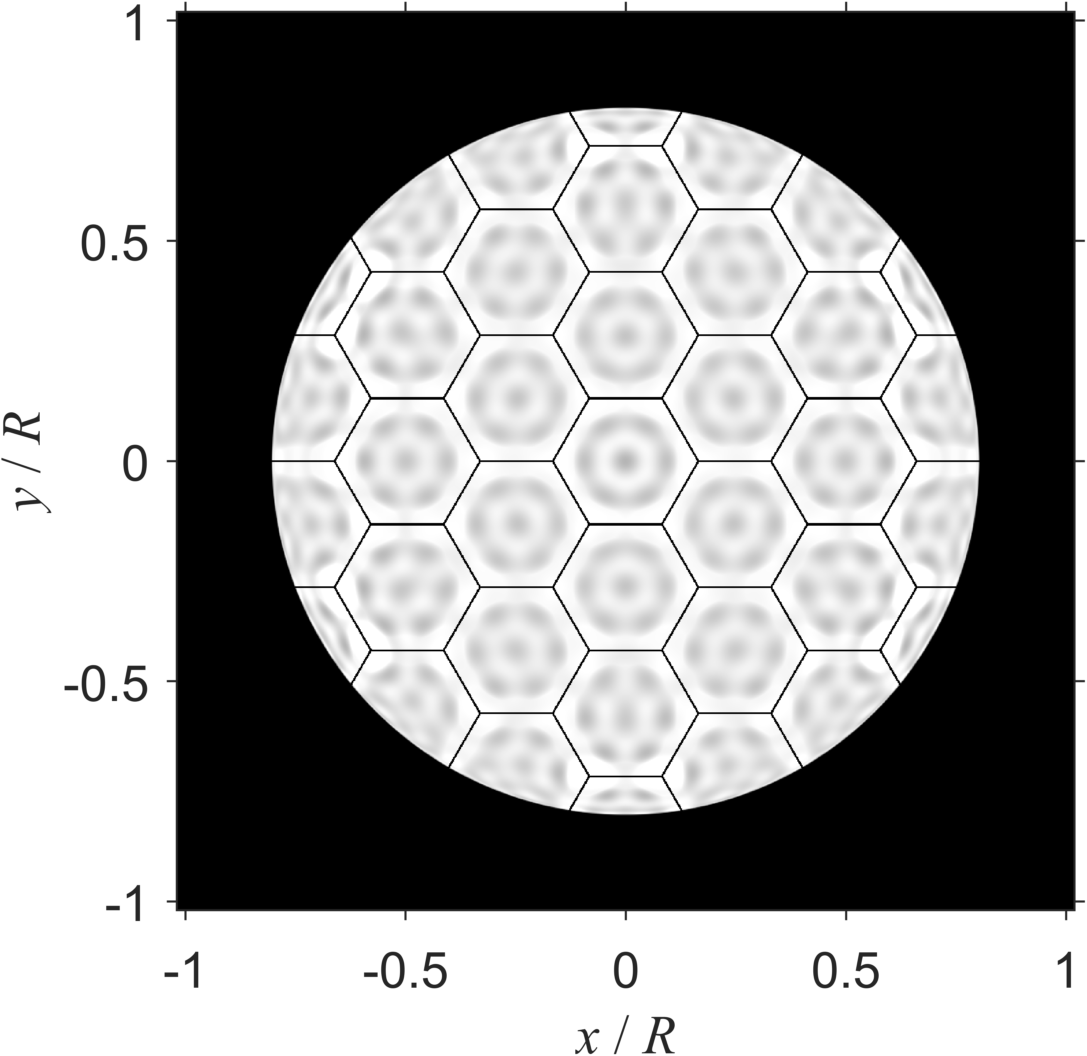}
    \includegraphics[height=5.45cm,trim={0.4cm 0.4cm 0 0},clip]{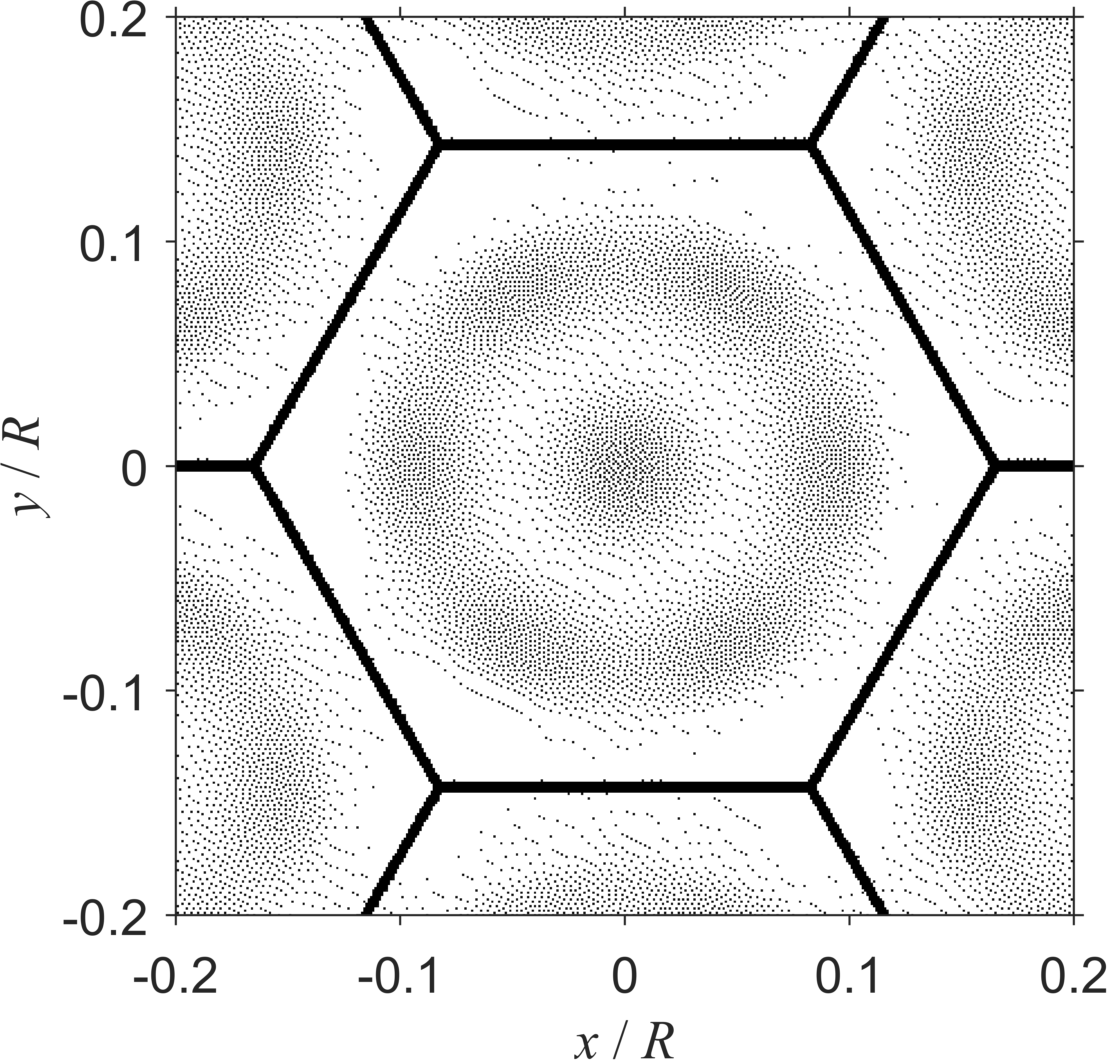}
    \caption{Apodizer design for an off-axis, segmented telescope with 37 hexagonal segments. \textit{Left:} The telescope aperture. \textit{Middle:} The desired amplitude pattern in the pupil to suppress starlight up to an angular separation of 20~$\lambda/D$ downstream of a vortex coronagraph. \textit{Right:} The microdot pattern designed to impart the desired amplitude pattern on the reflected field (zoomed in on the central segment). The diameter of the largest inscribed circle corresponds to $\sim$1700 microdots.}
    \label{fig:HCSTdesign}
\end{figure}

\begin{figure}[t!]
    \centering
    \includegraphics[height=4.5cm]{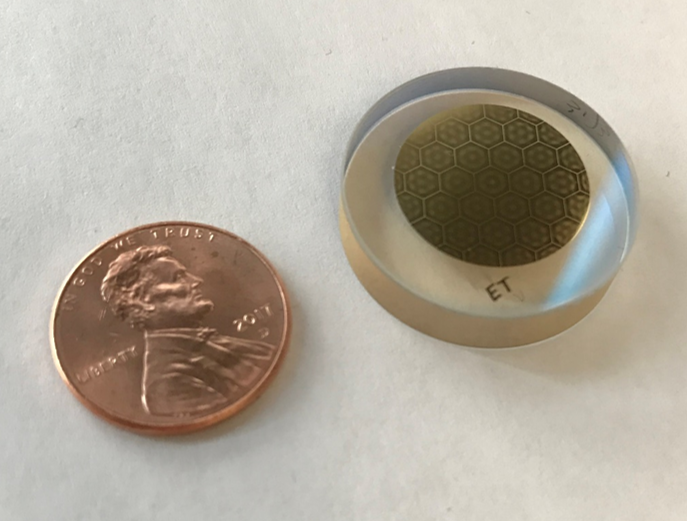}
    \includegraphics[height=4.5cm]{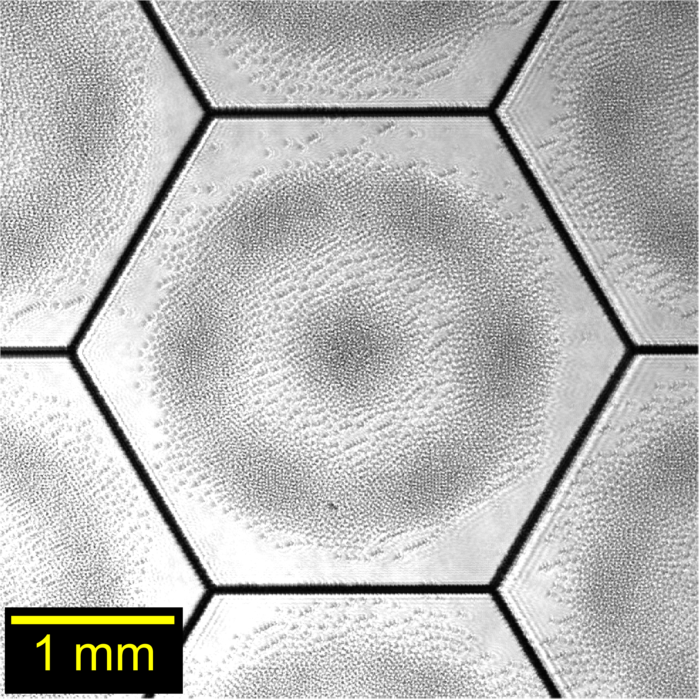}
    \includegraphics[height=4.5cm]{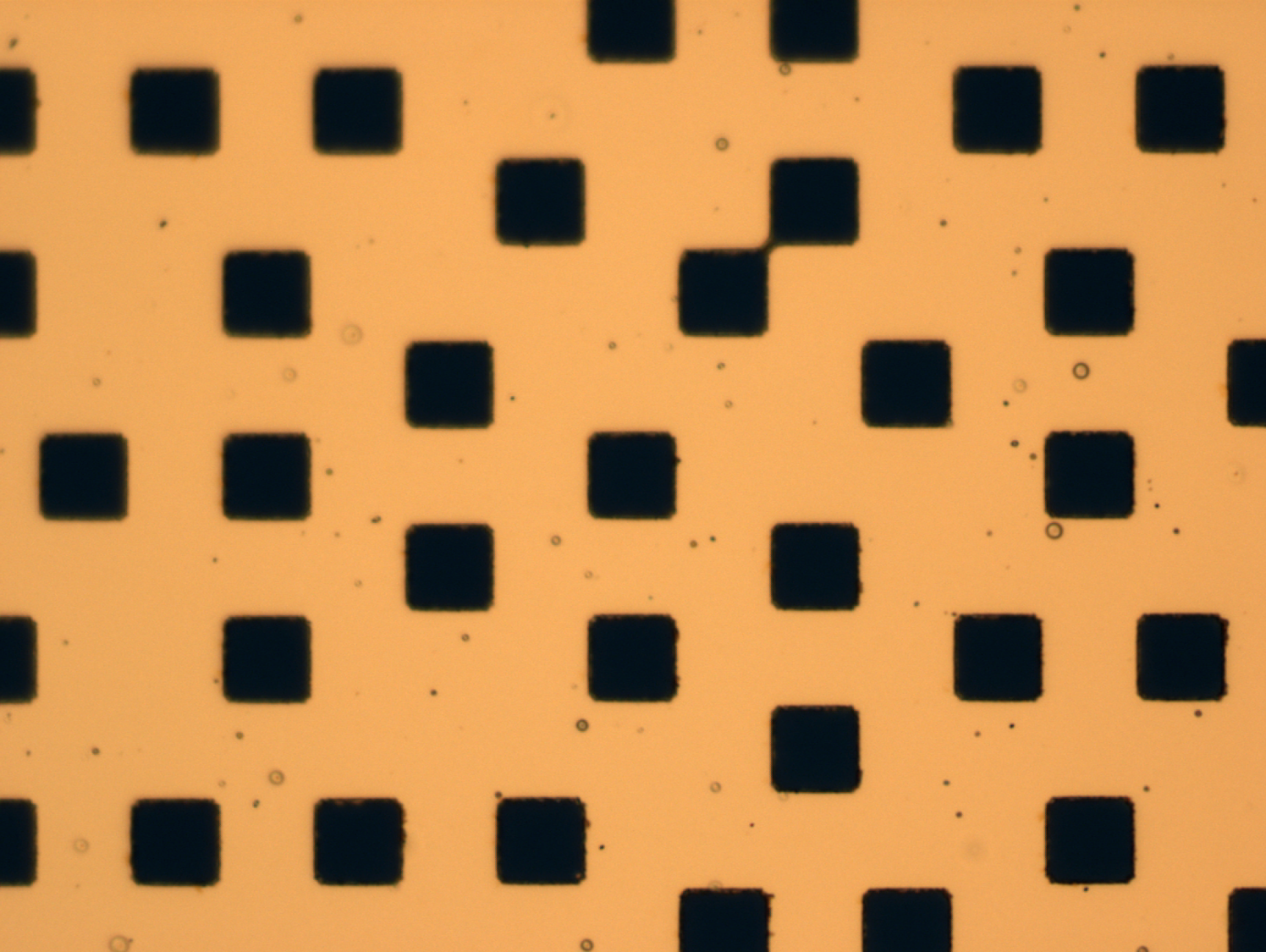}
    \caption{Gold microdot apodizer (10$\mu$m dot size) fabricated for testing on the High Contrast Spectroscopy Testbed (HCST) in Caltech's Exoplanet Technology (ET) lab.  \textit{Left:} A photograph of the sample with a U.S. one-cent coin for scale. \textit{Middle:} Zoomed in image on a single segment showing the manufactured microdot pattern. \textit{Right:} A microscope image of the square 10~$\mu$m microdots that make up the apodizer. }
    \label{fig:HCSTmicrodotapodizers}
\end{figure}

\section{Design and characterization of a microdot apodizer for a vortex coronagraph on a segmented telescope}

The HabEx and LUVOIR mission concepts may have a segmented primary mirror, such as the example off-axis, segmented telescope aperture with 37 hexagonal segments shown in Fig. \ref{fig:HCSTdesign}, \textit{left}. In that case, the gaps between mirror segments will cause unwanted starlight to diffract away from the star's position and possibly prevent the detection of new planets. Such diffraction may be suppressed by specially designed apodizers in the coronagraph instrument. For instance, the field amplitude pattern shown in Fig. \ref{fig:HCSTdesign}, \textit{middle} is tailored to minimize diffracted light up to an angular separation of 20~$\lambda/D$ downstream of a vortex coronagraph on an off-axis, segmented telescope \cite{Ruane2018_JATIS}. We fabricated a microdot apodizer designed to impart the desired pattern on the reflected field (see Fig. \ref{fig:HCSTdesign}, \textit{right}) using the Auxiliary Field Optimization technique\cite{Jewell2017}. We intend to incorporate this mask in an end-to-end demonstration of a coronagraph for exoplanet imaging with segmented aperture telescope on the High Contrast Spectroscopy Testbed (HCST) at Caltech's Exoplanet Technology (ET) lab. 

The fabricated apodizer mask was a $\sim$400~nm thick layer of evaporated gold in a 17~mm diameter region with $\sim$1700  square 10~$\mu$m microdots across a linear dimension. The gold region was made slightly elliptical to account for the 6$^\circ$ angle of incidence on the reflective apodizer on HCST. The substrate was a 6~mm thick slab of fused silica 25.4~mm in diameter with a surface flatness of $\lambda/20$ and AR coating for 650-1000~nm (Newport 10BW40-30AR.16). By design, the apodizer diameter is 83\% of the full pupil diameter (measured flat-to-flat). The individual segments were 2.9~mm across (flat-to-flat) with a 0.5~mm gap between them. A thin sub-layer of chrome was deposited in the manufacturing process prior to the evaporated gold to assist adhesion.

\begin{figure}[t!]
    \centering
    \includegraphics[height=6cm]{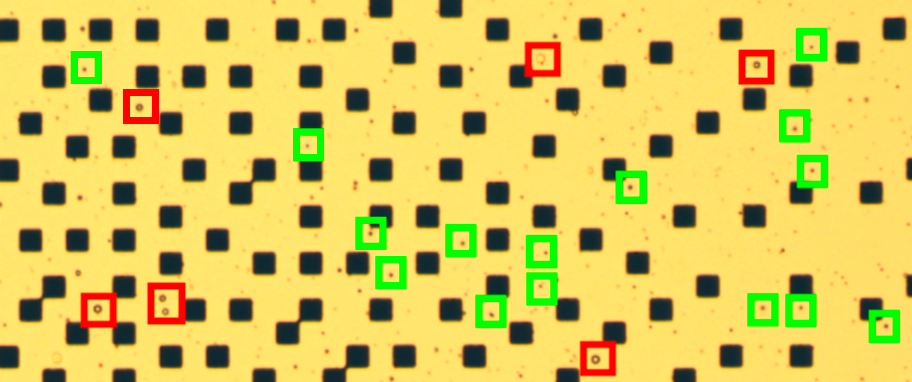}
    \caption{Microscope image of a 0.5$\times$0.2~mm section of the apodizer. The square dots are 10$\times$10~$\mu$m voids in the metallic layer where the AR coated substrate is exposed. The red squares show some of the ``type A'' defects, which are on average 0.25~$\mu$m in diameter, while the green squares show some of the ``type B'' defects, which are on average 0.13~$\mu$m in diameter}
    \label{fig:GoldApodizerDefects}
\end{figure}

Two different types of defects were found in the layer of evaporated gold (see Fig. \ref{fig:GoldApodizerDefects}). The first type (label as ``type A'') of defects are on average 0.25~$\mu$m in diameter and have the appearance of a semi-reflective circular spot. They occur with a density of approximately 0.002 defects/$\mu m^2$. ``Type B'' are a smaller with an average diameter half that of the ``type A'' defects ($\sim$0.13 $\mu m$), but are ten times as frequent (0.02 defects/$\mu m^2$). Type B defects are much darker and appear to be small pits in the gold layer. The effect of these defects would be on the performance of the coronagraph will be determined through end-to-end testing on the HCST.

\begin{figure}[h]
    \centering
    \includegraphics[height=5.7cm]{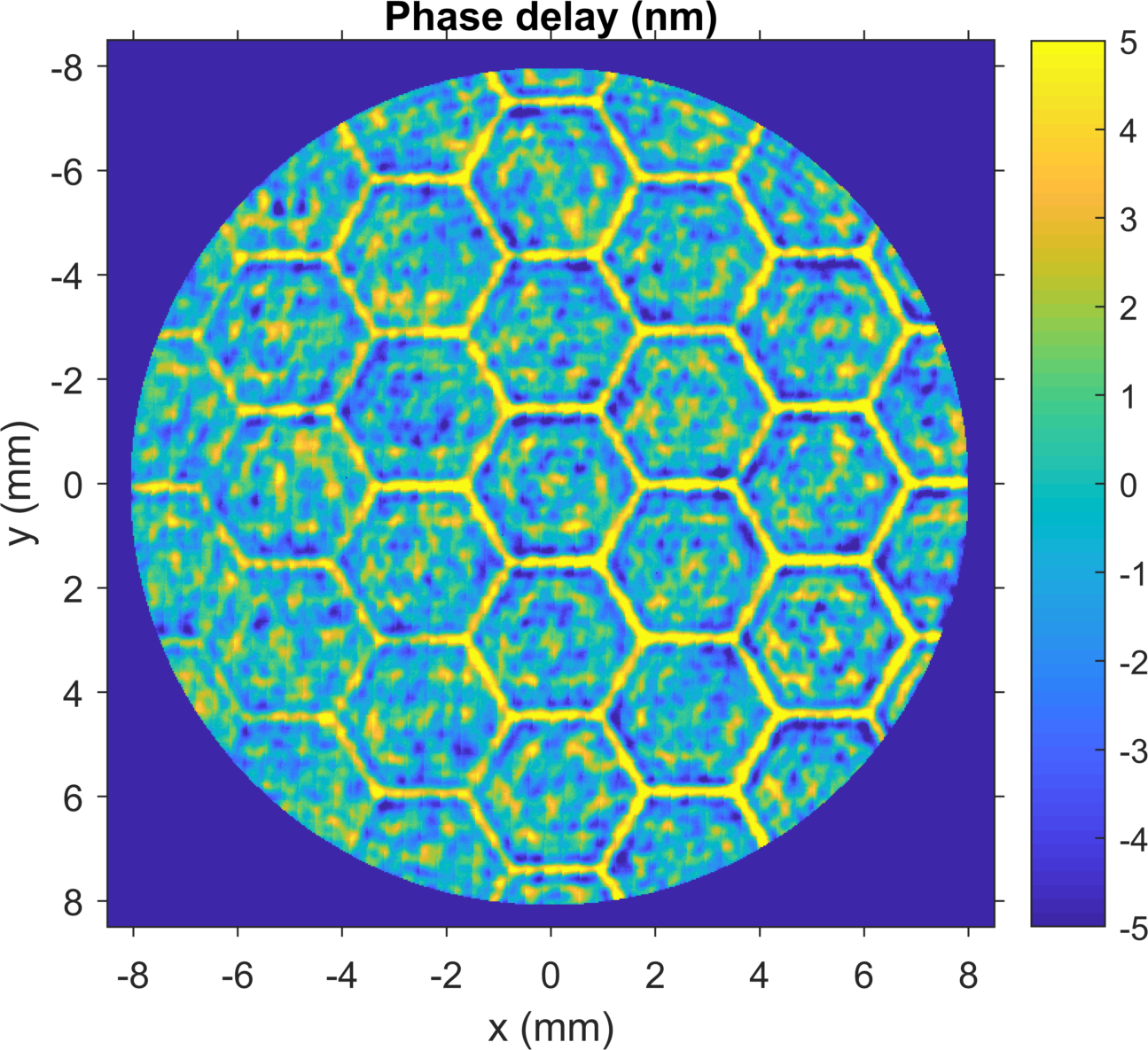}
    \caption{Wavefront error (in nm) imparted on the beam reflected from the apodizer shown in Fig. \ref{fig:HCSTmicrodotapodizers} measured with laser interferometry at $\lambda$=633~nm (after removing low-order Zernike polynomials). The hexagons are 2.93~mm flat-to-flat and the gaps between segments are $\sim$0.5~mm. Mid-spatial frequency aberrations potentially correlated with the local fill factor are only a few nanometers peak to valley.}
    \label{fig:HSCTzygo}
\end{figure}

Since the apodizer was intended for use around $\lambda$=800~nm, we tuned the thickness of the gold layer to be $\sim$400~nm to ensure that the fields reflected from the gold and substrate do not destructively interfere, the local reflectance closely matches the fill factor, and the reflected phase is as flat as possible. Figure \ref{fig:HSCTzygo} shows the phase of the reflected beam measured with laser interferometry. After removing low-order Zernike polynomials, the phase shifts that may be correlated with the local fill factor were small (only a few nanometers peak to valley). Larger phase delays were measured along the gaps where the reflectance is very low due to the AR coating. 

The apodizer was also designed to avoid enhanced transmission from periodic patterning of microdots by ensuring that the local transmission remains higher than 50\% everywhere on the sample. Although the darkest regions approach that level, periodicity isn't prevalent because the dark regions were localized. Overall, the reflected amplitude and phase from the apodizer closely matches the design as measured in the laboratory. 

\section{Conclusions and Future Work}

We have presented experiments performed to characterize microdot apodizers to be used in the discovery of exoplanets via imaging with future segmented aperture space telescopes. A test sample was used to investigate the relationship between the local fill factor of microdots and the measured wavefronts reflected and transmitted through the apodizer. We showed that analytical interference models are not ideal for describing the reflected amplitude and phase as a function of wavelength. Rather, it is likely that the incident field was interacting with electronic resonances at the surface of the metal layer. These effects were amplified as the dot size approaches the wavelength and as periodic patterns appeared in the microdot arrays near fill factors of $\sim$50\%. Future work will apply finite difference time domain (FDTD) and rigorous coupled wave analysis (RCWA) techniques to model these effects. 

Simultaneously, experiments are underway to demonstrate the apodizer fabricated for a segmented aperture telescope as part of an end-to-end coronagraph instrument. For this purpose, a second generation apodizer was designed for a vortex coronagraph, which will be tested this year on the High Contrast Spectroscopy Testbed (HCST) at Caltech's Exoplanet Technology (ET) lab. The choice of thickness of the metallic layer, dot size, and materials mitigated phase effects that correlate with the local fill factor (few nanometers peak to valley). The result is an apodizer mask with well tuned reflectance to assist in the control unwanted diffraction of starlight in space-based coronagraphs.  

\acknowledgments     
G. Ruane is supported by an NSF Astronomy and Astrophysics Postdoctoral Fellowship under award AST-1602444. This work was supported by the Exoplanet Exploration Program (ExEP), Jet Propulsion Laboratory, California Institute of Technology, under contract to NASA.

%%%%%%%%%%%%%%%%%%%%%%%%%%%%%%%%%%%%%%%%%%%%%%%%%%%%%%%%%%%%%
%%%%% References %%%%%

\bibliography{RuaneLibrary}   %>>>> bibliography data in report.bib
\bibliographystyle{spiebib}   %>>>> makes bibtex use spiebib.bst

\end{document}